\documentclass[onecolumn,twoside,dvipdfm]{IEEEtran}
\usepackage{bbm}
\usepackage{amsmath,amsthm,amsfonts, amssymb,mathdots}
\usepackage{mathrsfs}

 \newtheorem{thm}{Theorem} [section]

\newtheorem{definition}[thm]{Definition}
\newtheorem{lem}[thm]{Lemma}

 \newtheorem{conj}[thm]{Conjecture}
 \newtheorem{prob}[thm]{Problem}

\newcommand{\mP}{\mathcal{P}}

\def\f#1{{\mathbb{F}}_{#1}}
\newcommand{\bP}{\mathbb{P}}
\newcommand{\bbsm}{ \begin{smallmatrix}} 
\newcommand{\besm}{\end{smallmatrix}}

\begin{document}

\title{Deep Holes of Projective Reed-Solomon Codes\thanks{
 The research of Jun Zhang was supported by the National Natural Science Foundation of China under Grant No. 11971321, and by Scientific Research Project of Beijing Municipal Education Commission under Grant No. KM201710028001. Jun Zhang is supported by Chinese Scholarship Council for his visit at the University of Oklahoma, USA.  The research of Daqing Wan was supported by National Science Foundation.  The research of Krishna Kaipa is supported by  the Science and Engineering Research Board, Government of India, under Project EMR/2016/005578.
}}

\author{Jun Zhang\thanks{Jun Zhang is with the School of Mathematical Sciences, Capital Normal University, Beijing 100048, China. Email: junz@cnu.edu.cn},
\and
Daqing Wan\thanks{Daqing Wan is with the Department of Mathematics, University of California, Irvine, CA 92697, USA. Email: dwan@math.uci.edu},
\and
Krishna Kaipa\thanks{Krishna Kaipa is with the Department of Mathematics, Indian Institute of Science Education and Research, Pune, Maharashtra, 411008 India. Email: kaipa@iiserpune.ac.in}
}

\date{}
\maketitle

\begin{abstract}
Projective Reed-Solomon (PRS) codes are Reed-Solomon codes of the maximum possible length $q+1$.  The classification of deep holes --received words with maximum possible error distance-- for PRS codes is an important and difficult problem. In this paper, we use algebraic methods to explicitly construct three classes of deep holes for PRS codes. We show that these three classes completely classify all deep holes of PRS codes with redundancy  four. Previously, the deep hole classification was only known for PRS codes with redundancy at most three by the work  \cite{Kaipa17}.
\end{abstract}
\IEEEpeerreviewmaketitle

\section{Introduction} \label{introduction}
Let $\f{q}$ denote the finite field of size $q$ and characteristic a prime number $p$. Let $\f{q}^n$ denote the vector space of row-vectors  or words $ {x}=(x_1,x_2,\cdots,x_n)$.  The Hamming metric on  $\f{q}^n$  is  the metric obtained by defining the distance between two words $x$ and $y$ to be the number of coordinates in which $x$ and $y$ differ:
\[ d(x,y)= | \{ i \,|\, x_i \neq y_i\}|. \]
A  \emph{linear $[n,k]$ code} $C$ is a $k$-dimensional linear subspace of $\f{q}^n$ with the induced metric. The \emph{minimum distance} $d(C)$ of $C$ is 
\[ d(C)=\min\{d(x,y)  \,|\,  \text{ $x$ and $y$ are distinct elements of $C$}\}.\]
The \emph{error distance} of any word $u\in\f{q}^n$ to $C$ is defined to be
\[ d(u,C)=\min\{d(u,v)\,|\,v\in C\}. \]
The maximum error distance
\[
   \rho(C)=\max\{d(u,\, C)\,|\,u\in \f{q}^n\},
\]
is called the \emph{covering radius} of $C$, and the words achieving maximum error distance are called \emph{deep holes} of the code. The problem of determining the set of deep holes of a code, or of deciding whether a given word is a deep hole, are in general hard problems. These problems are also important from the perspective of  the decoding problem for the code. \\

When $C$ is a Reed-Solomon code, the problem of determining the deep-holes of $C$ is an interesting and difficult combinatorial problem. This problem has received significant attention in recent literature, for example in the works \cite{CMP11} \cite{CM07} and \cite{Kaipa17} \cite{KW15} \cite{LZ15} \cite{ZW16} \cite{ZCL16}.  In this work, by a Reed-Solomon code we mean the following code:
\begin{definition}
Let  $D= (x_1, \dots, x_n)$ be an ordered set of  $n$ distinct elements of $\f{q} \cup \infty$. The Reed-Solomon code $RS(D,k)$ of length $n$, dimension $k$ and evaluation set $D$ is  the code:
\[
    RS(D,k)=\{(f(x_1),\ldots,f(x_n))\in \f{q}^n \,\, | \,\,  f(X)\in \f{q}[X], \deg(f)\leq k-1\}. \]
\end{definition}
Here $f(\infty)$ is taken to be the coefficient of $X^{k-1}$ in $f(X)$, and the parameters $n$ and $k$ satisfy  $1 \leq k \leq n \leq q+1$.
Generalized Reed-Solomon (GRS) codes are obtained by applying a diagonal Hamming isometry to a Reed-Solomon code:  in other words, any GRS code is of the form  $C'=\{c M :  c \in C\}$ where $C$ is a $[n,k]$ Reed-Solomon code and $M$ is an invertible  $n \times n$ diagonal matrix over $\f{q}$.  Clearly, the set of deep  holes of  $C'$ is $\{x M \,|\,  x \text{ is a deep hole of $C$}\}$. Therefore,  for the problem of determining the deep holes of GRS codes, it suffices to only treat Reed-Solomon codes.\\

When $D = \f{q} \cup \infty$, the Reed-Solomon codes $RS(D,k)$ are called \emph{projective Reed-Solomon} codes and will be simply denoted as $PRS(k)$. These codes are also known in literature as  doubly-extended Reed-Solomon codes.
 While the covering radius of $[n,k]$ Reed-Solomon codes of length $n <q+1$ is known to be $n-k$, the situation with PRS codes is different. 
 For $k \in \{1,q,q+1\}$, the covering radius is again $n-k$ and the deep holes of $PRS(k)$ are easily determined. For $k=q-1$, the covering radius is $n-k-1$ and the deep holes of $PRS(k)$ are known  (see  \S \ref{PRSCode} for these facts).  But for  $2 \leq k \leq q-2$,  the covering radius of $PRS(k)$ is only known conjecturally:
\begin{conj}  \label{cover}
For $2 \leq k \leq q-2$, the covering radius of $PRS(k)$ is: 
\[ \begin{cases} 
 q-k+1 &\text{if $q$ is even and $k \in\{2, q-2\}$} \\
q-k  &\text{otherwise.}
\end{cases}\]
\end{conj}
This conjecture is equivalent to a well-known conjecture in finite geometry (see Conjecture \ref{cover_rnc} in \S \ref{sec_cov_PRS}). Conjecture \ref{cover} has been 
shown to be true for several values of $k$ for example $k\geqslant \lfloor (q-1)/2 \rfloor$.
This brings us to our main problem:
\begin{prob}\label{prob}
Determine the set of  deep holes  $PRS(k)$  for those $k$ for which Conjecture \ref{cover} is true.
\end{prob}
For  $k\in \{q-1, q-2\}$, the problem has an easy solution as given in \cite{Kaipa17}  (see \S \ref{PRSCode} for details).  For  $2 \leqslant k \leqslant q-3$, the problem is  difficult and wide open.  In~\cite{ZW16}, the following  classes  of deep holes of $PRS(k)$ were identified:
\begin{thm}[~\cite{ZW16}]\label{deephole:degreek}
Let   $2\leq k\leq q-3$ and  suppose $\rho(PRS(k))=q-k$.
Let   $D = (\alpha_1, \dots, \alpha_q, \infty)$ be the ordered evaluation set for $PRS(k)$.	The $q$  words 
	\[ \{ ( \alpha_1^k, \alpha_2^k, \dots, \alpha_q^k, a)\,: \, a \in \f{q}\},\] are distinct deep hole classes of $PRS(k)$.
\end{thm}
The term deep hole \emph{class} is defined in the next section. The automorphism group of a linear code also acts on the set of its deep holes. Using this fact, we show in \S \ref{sec_aut}  that the group $PGL_2(\f{q})$ acts on the set of deep hole classes of $PRS(k)$. The orbits under $PGL_2(\f{q})$ of the $q$ deep hole classes given in Theorem \ref{deephole:degreek} above, give us new classes of deep holes of $PRS(k)$. This is the first result of this paper:
\begin{thm}
\label{thm:Rotationsofdegk}
Let $2\leq k\leq q-3$ and suppose $\rho(PRS(k))=q-k$. The set of words 
\[\{ 
	(\tfrac{1}{\alpha_1-\alpha_i},\cdots,\tfrac{1}{\alpha_{i-1}-\alpha_i}, a,\tfrac{1}{\alpha_{i+1}-\alpha_i},\cdots, \tfrac{1}{\alpha_q-\alpha_i}, 0)
\, : \, 1 \leq i \leq q, \, a \in \f{q} \},	\]
represent  $q^2$ distinct classes of deep holes of $PRS(k)$. These classes are distinct from the $q$  classes of Theorem \ref{deephole:degreek}.
\end{thm}
We also show that the $q^2+q$ deep hole classes of Theorems \ref{deephole:degreek} and \ref{thm:Rotationsofdegk}  taken together  have a nice geometric interpretation in terms of  the tangent lines to the degree $(q-k)$ normal rational curve in $\bP^{q-k}(\f{q})$. \\

The second result of this paper is a new class of deep holes of $PRS(k)$:
\begin{thm} \label{thm:main_irr2}
Let $2\leq k\leq q-3$ and suppose $\rho(PRS(k))=q-k$. The  words $(g(\alpha_1), g(\alpha_2), \dots, g(\alpha_q),0)$	
as $g$ runs over the $(q+1)q(q-1)/2$ rational functions of the form $a(X)/b(X)$ with $b(X)$ a monic  irreducible polynomial of degree $2$, and $a(X)$ a nonzero monic polynomial of degree at most $1$, represent  $(q+1)q(q-1)/2$ distinct classes of deep holes of $PRS(k)$.
These classes are distinct from the $q^2+q$  classes of Theorems \ref{deephole:degreek} and \ref{thm:Rotationsofdegk}.
\end{thm}
We also show that this construction has a geometric interpretation in terms of   the degree $(q-k)$  normal rational curve in $\bP^{q-k}(\f{q^2})$ over a quadratic field extension $\f{q^2}$ of $\f{q}$.\\

The third result of this paper is the  complete classification of deep holes of $PRS(k)$ for $k=q-3$:
\begin{thm}\label{main:k=q-3} The total number of deep hole classes of $PRS(q-3)$ is  $q(q+1)^2/2$.   These are given by  the $q$ deep hole classes of Theorem~\ref{deephole:degreek}, $q^2$ classes of Theorem~\ref{thm:Rotationsofdegk}, and $(q+1)q(q-1)/2$ classes of Theorem~\ref{thm:main_irr2}.
\end{thm}
The rest of this paper is organized as follows. In \S  \ref{sec_new_deepholes}
we prove Theorems~\ref{thm:Rotationsofdegk} and~\ref{thm:main_irr2}. The necessary tools and background are covered in \S \ref{sec_aut} and 
 and \S \ref{sec_cov_PRS}. We prove Theorem \ref{main:k=q-3}
 in \S \ref{section:k=q-3}.  The purpose of \S \ref{sec_cov_MDS}
 is to highlight the fact that  the assertion which sometimes appears in literature-- that MDS codes of minimum distance $d$ have covering radius $(d-1)$ or $(d-2)$ -- has no known correct proof. This section is independent of the rest of the paper.

\section{Projective Reed-Solomon codes and their Covering Radii}\label{PRSCode}
We begin with some notation for PRS codes. We use the term words for row vectors, and a vector (of $\f{q}^m$ for some $m$) 
will mean a column vector.  For any integer $1 \leq k \leq q+1$  and $\alpha \in \f{q} \cup \infty$, we define vectors
\begin{equation} \label{eq:cdef}
c_k(\alpha)=\begin{cases}
(1,\alpha,\alpha^2,\cdots,\alpha^{k-1})^T & \mbox{ if $\alpha\in\f{q}$,}\\
(0,0,\cdots, 0, 1)^T & \mbox{ if $\alpha=\infty$}.
\end{cases}
\end{equation}
For $k=1$, it is understood that $c_k(\alpha)=1$ for all $\alpha \in \f{q} \cup \infty$. Let $\alpha_1, \dots, \alpha_q$ be  a fixed  ordering of $\f{q}$, and let $\alpha_{q+1}= \infty$.
The matrix $G_k$ defined as:
 \begin{equation} \label{eq:G_k} G_k=[c_k(\alpha_1) | \dots | c_k(\alpha_q) | c_k(\alpha_{q+1}) ], \end{equation}
is a generator matrix for $PRS(k)$: for a message word $(a_0, \dots, a_{k-1})$ the codeword $(a_0, \dots,a_{k-1}) G_k$ is the evaluation of the polynomial $a_0 + a_1 X + \dots + a_{k-1} X^{k-1}$ at  the ordered set of points $\alpha_1, \dots, \alpha_{q+1}$.
We recall that for a polynomial  $f(X)$ of degree at most $k-1$, the value of $f(\infty)$ is taken to be the coefficient of $X^{k-1}$ in $f(X)$.  Any $k \times k$ minor of $G_k$ is a Vandermonde determinant  and hence is nonzero (here we use the fact that $k \leq q+1$).
It is well known that a linear $[n,k,d]$ code $C$ satisfies $d \leq n-k+1$ (the Singleton bound), and  equality holds in this bound if and only if every $k \times k$ minor of a generator matrix of $C$ is nonzero. Such a code is called a maximum-distance-separable code (MDS code). Therefore $PRS(k)$ is always an  MDS code.
Using the following well-known identity about sum of powers of elements in $\f{q}$:
	\begin{equation} \label{eq:sumofpowers}
		\sum_{\alpha\in\f{q}^{\times}} \alpha^i=\begin{cases}
		0 & \mbox{if $q-1 \nmid i$}\\
		-1 & \mbox{if $q-1 \mid i$,}
		\end{cases} \end{equation}
it follows that  for $1 \leq k \leq q$,  the product $G_k G^T_{q+1-k}$ is the $k \times (q+1-k)$ zero matrix. Thus  $G_{q+1-k}$ is a parity check matrix for $PRS(k)$, or equivalently  $PRS(q+1-k)$ is the dual code to $PRS(k)$.  \\

For a linear code $[n,k]$ code $C$, and a  received  word $u \in\f{q}^n$, the word $v=a u + c$ where $c \in C$ and $a \in \f{q}^{\times}$ has the same error distance as $u$.  We recall that for a vector space $V$, the projective space  $\bP(V)$ denotes the set of equivalence classes of $V \setminus \{0\}$ in which two nonzero vectors are equivalent if and only if they generate the same one dimensional subspace of $V$. 
\begin{definition} \label{equiv}
For a linear code $[n,k]$ code $C$, we will say that received words $u,v \in \f{q}^n$ are equivalent if $v=a u + c$ for some  $c \in C$ and $a \in \f{q}^{\times}$. In particular  non-codewords $u, v$ are equivalent if and only if they represent the same element of the projective space  $\bP(\f{q}^n/C)$. Here  $\f{q}^n/C$ is the quotient vector space of cosets of $C$ in $\f{q}^n$.
The term \emph{deep hole class} will refer to the class of a deep hole of $C$ in $\bP(\f{q}^n/C)$. 
\end{definition}
 
We recall the following well-known characterization of covering radius  of a linear code in terms of a parity check matrix.
   \begin{lem}\label{coveringradiusfromparitycheck}
   Let $C$ be a $[n,k]$ linear code with  parity check matrix $H$. The error distance $d(u,C)$ of a received word  $u\in \f{q}^n$	equals the least number $j$ such that  the syndrome $\mathrm{syn}(u)=Hu^T$ can be expressed as a linear combination of $j$  columns of $H$.  In particular, the covering radius $\rho(C)$ is the least integer $j$ such that any vector in $\f{q}^{n-k}$ can be expressed as a linear combination of some $j$ columns of $H$. 	The word $u$ is a deep hole of $C$ if and only if  $\mathrm{syn}(u)$ can not be written as a linear combination of any $\rho(C)-1$ columns of $H$.
   	   \end{lem}

 \begin{definition} \label{projsyn}
Let $C$ be a $[n,k]$ linear code.  The term \emph{projective syndrome} of a non-codeword $u$ will refer to the element of $\bP(\f{q}^{n-k})$ represented by $\mathrm{syn}(u)$. The  map $u \mapsto \mathrm{syn}(u)$ induces a bijective correspondence 
   \begin{equation} \label{eq:projsyn}  \mathrm{syn} : \bP(\f{q}^n/C) \to \bP(\f{q}^{n-k}) \end{equation}
   from the set $\bP(\f{q}^n/C)$ of equivalence classes of non-codewords  to the set of projective syndromes $\bP(\f{q}^{n-k})$.
\end{definition}
We use the notation $\mathbb S(k)$ for the subset of $\bP^{q-k}(\f{q})=\bP(\f{q}^{q+1-k})$ consisting of the projective syndromes of deep hole classes of $PRS(k)$. We note that the Problem \ref{prob} is equivalent to determining the subset  $\mathbb S(k) \subset \bP^{q-k}(\f{q})$. 
In the literature  on deep holes of RS codes, deep holes are often described by \emph{generating polynomials}. This can be adapted to PRS codes and is  closely related to our description in terms of projective syndromes: Let $\mP^{q-k}(\f{q})$ denote the set of  polynomials of degree at most $(q-1)$ which are monic and for which the coefficient of $1, X, \dots, X^{k-2}$ is zero. The number of such polynomials is $1+q+\dots+q^{q-k}$. To each such polynomial, we associate the word $u=(u(\alpha_1), \dots, u(\alpha_q),0) \in \f{q}^{q+1}$. 
If $u(X) \neq v(X)$, we claim $u$ and $v$ represent different equivalence classes: if $v= a u + c$ for some  $a \in \f{q}^{\times}$  and $c \in C$, then there is a polynomial $f(X)$ of degree at most $k-1$  (representing the codeword $c$) such that $v(X)-a u(X) -f(X)$ has $q$ roots, but degree at most $q-1$. This forces $v(X)-a u(X) =f(X)$.
We note that 
\[ 0= v_{q+1}-a u_{q+1} = c_{q+1} = f(\infty).\]
Since $f(\infty)$ is the coefficient of $X^{k-1}$ in $f(X)$,   it follows that  deg$(f) \leq k-2$.  Combining this with the fact that the coefficients of $1, X, \dots, X^{k-2}$ in $u(X)$ and $v(X)$ are zero, forces $f(X)= v(X)- a u(X) =0$. Since $u(X), v(X)$ are monic we get $a=1$ and hence $u(X)=v(X)$. The deep hole class of $u$ is said to be \emph{generated} by the polynomial $u(X)$. The relation between the projective syndrome syn$(u)$ and the polynomial $u(X)$ is very simple: if $u(X)= \sum_{i=1}^{q-k+1} a_i X^{q-i}$ generates the word $u$, then the projective syndrome syn$(u)=(a_1:a_2: \dots:a_{q-k+1})$.  This easily follows from the formula syn$(u)=G_{q+1-k} u^T$ together with the identity \eqref{eq:sumofpowers}. For example the projective syndromes of the $q$ words of Theorem \ref{deephole:degreek} are: 
\begin{equation}  \label{eq:syn_degreek}
\text{syn}( \alpha_1^k, \alpha_2^k, \dots, \alpha_q^k, a)= (0: \dots:0:1:-a),
\end{equation}
 and the corresponding generating polynomials are $X^k- a X^{k-1}$.

\subsection{Covering radius of PRS codes} \label{sec_cov_PRS}
We first discuss the possible values of covering radius for PRS codes. As mentioned above, the matrix $G_{q+1-k}$ is a parity check matrix for the code $PRS(k)$. Since  $G_{q+1-k}$ has full rank, it follows from Lemma \ref{coveringradiusfromparitycheck}, that $\rho(PRS(k)) \leq q+1-k$. Next, we  show that  $\rho(PRS(k))=q+1-k$ for  $k \in \{1,q,q+1\}$, 
and $\rho(PRS(k))=q-k$ for $k=q-1$. We also determine the set of deep holes in each case:
\begin{itemize}
\item For $k=q+1$, we have $\rho(PRS(k))=0$ and hence every word is a deep hole: this is because $\rho(PRS(k)) \leq q+1-k=0$.
\item For $k= q$, we have $\rho(PRS(k)) =1$, and hence every non-codeword is a deep hole: here $\rho(PRS(k)) \leq q+1-k=1$, and $\rho(PRS(k)) \neq 0$ because a linear $[n,k]$ code $C$ satisfies $\rho(C)=0$ if and only if $n=k$.
\item  For $k= q-1$, we have $\rho(PRS(k)) =1$, and hence every non-codeword is a deep hole: here any syndrome in $\f{q}^2$ is proportional to one of  the $(q+1)$ columns of the parity check matrix $G_2$, and hence  $\rho(PRS(k)) =1$. 
\item  For $k=1$,  we have $\rho(PRS(k)) =q$:  here the codewords  are $\{(a, \dots, a) : a \in \f{q}\}$ and hence the maximum possible distance of a received word from the code is $q$. The deep holes are those received words of length $q+1$ which have the maximum possible number (namely $q$) of  distinct coordinates.\\
\end{itemize}
On the other hand, for $2 \leq k \leq q-2$ it is also known that $\rho(PRS(k)) \geq q-k$: for the  word $u=(\alpha_1^k, \dots, \alpha_q^k,0)$,  it can be shown that $d(u, PRS(k))=q-k$. A quick proof is as follows. The distance of $u$ from a codeword represented by a polynomial $f(X)$ of degree at most $k-1$, is at least $q-k$ because the polynomial $X^k - f(X)$ can have at most $k$ roots. On the other hand for   $f(X)=X^k-(X-x_1)(X-x_2) \dots(X-x_k)$ where $x_1, \dots, x_k$ are  $k$ distinct elements of $\f{q}$ which add up to $0$ (this is always possible, see  \cite{GK98}, \cite{ZW16}), the distance of $u$ from the  codeword represented by $f(X)$ is exactly $q-k$. For $2 \leq k \leq q-2$,  we have $\rho(PRS(k))= q+1-k$ if there exists a vector $v \in \f{q}^{q+1-k}$ which cannot be expressed as a linear combination of any $q-k$ columns of $G_{q+1-k}$ (by Lemma \ref{coveringradiusfromparitycheck}). If no such vector  exists, then  
$\rho(PRS(k)) \leq q-k$ and hence $\rho(PRS(k)) =q-k$.  As mentioned in the introduction, a $[n,k]$ linear code is MDS if and only if every $k\times k$ minor of a generator matrix of the code is nonzero. Since this property holds for $G_{q+1-k}$, we can rewrite the above characterization of $\rho(PRS(k))$ in the  following way (originally due to   D\"{u}r  (1994)):

\begin{lem}  \label{lem_cover} \cite{Arne94}  For $2 \leq k \leq q-2$, we have $\rho(PRS(k))= q-k+1$ or $\rho(PRS(k))= q-k$ according as whether or not there exists a vector $v \in \f{q}^{q+1-k}$  such that the $(q+1-k) \times (q+2)$ matrix $[G_{q+1-k} | v]$ generates an MDS code,  i.e. whether or not there exists a $[q+2,q+1-k]$ MDS code extending $PRS(q+1-k)$ by one coordinate. 
\end{lem}

 In finite geometry an (ordered) $n$-arc in projective space $\bP^{m-1}(\f{q})$ is an ordered set of $n$ points of $\bP^{m-1}(\f{q})$ represented  by vectors $v_1, \dots, v_n \in \f{q}^m$ with the property that the $m \times n$ matrix $[v_1 | v_2 | \dots | v_n]$ generates a $[n,m]$ MDS code. The standard degree $(m-1)$ normal rational curve in $\bP^{m-1}(\f{q})$ is the image of the embedding 
 \[ \bP^1(\f{q}) \hookrightarrow \bP^{m-1}(\f{q}) \quad \text{given by} \quad  (x:y) \mapsto (x^{m-1}:x^{m-2}y: \dots:xy^{m-2}:y^{m-1}),\] or equivalently $t \mapsto c_m(t)$ where $t$ represents $(1:t)$ if $t \in \f{q}$, and $(0:1)$ if $t=\infty$. To keep the notation simple, we use the same symbol $c_m(t)$ for the class in $\bP^{m-1}(\f{q})$ of the vector $c_m(t) \in \f{q}^m$. 
 When $m \leq q$, these $(q+1)$ points form a $(q+1)$ arc in $\bP^{m-1}(\f{q})$ because they represent the $(q+1)$ columns of the matrix $G_m$. A $n$-arc in $\bP^{m-1}(\f{q})$ is said to be complete if it is not a subset of a $(n+1)$-arc in $\bP^{m-1}(\f{q})$. Therefore, Lemma \ref{lem_cover}  can be restated in finite geometry terms as: \\

\noindent \emph{Lemma \ref{lem_cover} restated:} For $2 \leq k \leq q-2$, we have $\rho(PRS(k))= q-k$ or $\rho(PRS(k))= q-k+1$ according as whether the $(q+1)$ points of the degree $(q-k)$ normal rational curve in $\bP^{q-k}(\f{q})$ form a complete arc or not.\\  A well-known  conjecture in finite geometry is: 
\begin{conj} \label{cover_rnc}
For $2 \leq k \leq q-2$, the $(q+1)$ points of the degree $(q-k)$ normal rational curve in $\bP^{q-k}(\f{q})$ form a complete arc 
except when $q$ is even and $k \in\{2, q-2\}$.
\end{conj}
We note that Conjecture \ref{cover} mentioned in the introduction is just a restatement of  Conjecture \ref{cover_rnc}. 
The conjecture is true if $k\geqslant \lfloor (q-1)/2 \rfloor$  from the work of Seroussi and Roth \cite{Ro_Se}. 
This was improved to  the range  $k \geq  6 \sqrt{q \ln q}  -2$  in \cite{Storme_complete}.   Also, Conjecture \ref{cover_rnc} is a special case  of the famous MDS conjecture which states that for $2 \leq k \leq q-1$, the maximum possible length of a $(q+1-k)$-dimensional MDS code is $q+1$ except when $q$ is even and $k \in \{2,q-2\}$.
Therefore,  Conjecture \ref{cover_rnc} is true for a value of $k$ if the MDS conjecture is true for the same $k$. Some  values of $k$ for which the 
 MDS conjecture is true are:
 \begin{enumerate}
 \item  \cite[Theorem 1.10]{Ball12}:        $3 \leq k \leq p \leq q-2$  
 \item   \cite{Thas1968}, \cite{Kaneta_Maruta}:     $q$ odd,    $q-\sqrt{q}/4 -9/4 < k  \leq q-2$.
\item \cite{Storme_Thas1993}:  $q$ even,  $  q-\sqrt{q}/2- 11/4 < k \leq q-4$.
\end{enumerate}
 Some other values of $k$ for which the Conjecture  \ref{cover_rnc} has been proved (for example $k \in\{2,3,4\}$ for $q$ odd)   can be found in \cite[\S4]{Kaipa17}.
\\

For $k=q-2$,  Problem \ref{prob} was solved in \cite{Kaipa17}:
\begin{thm}\label{main:k=q-2}
 Let $k=q-2$. \begin{itemize}
 \item If $q$ is even, then  $\rho(PRS(k))=3$ and $\mathbb S(k)=\{(0:1:0)\}$. In other words, there is exactly one deep hole class and its projective syndrome is $(0:1:0)$.
 \item If $q$ is odd, then   $\rho(PRS(k))=2$. There are $q^2$ deep hole classes, and $\mathbb S(k)$ consists of all points of $\bP^2(\f{q})$ other than  the $(q+1)$ points $\{c_3(t) : t \in \f{q} \cup \infty\}$. 
  \end{itemize}
  \end{thm}
Briefly, the problem of finding all $v \in \f{q}^3$ such that the matrix $[G_3 \,|\, v]$ generates a $[q+2,3]$ MDS code is easily seen to have no solution if $q$ is odd, and if $q$ is even then $v$ must be $(0,a,0)^T$ for $a \neq 0$. Thus by Lemma \ref{lem_cover}, $\rho(PRS(q-2))=2$ if $q$ is odd,  and $\rho(PRS(q-2))=3$ if $q$ is even. Next, by Lemma \ref{coveringradiusfromparitycheck}, $v \in \f{q}^3$ is the syndrome of a deep hole if and only if 
\begin{enumerate}
\item  Case when $q$ is odd: $v$ cannot be expressed as a linear combination of one column of $G_3(\f{q})$. In other words, the class of $v$ in $\bP^2(\f{q})$ consists of the $q^2$ points of $\bP^2(\f{q})$ other that $\{c_3(t) : t \in \f{q} \cup \infty\}$.
\item  Case when $q$ is even: the matrix $[G_3(\f{q}) \,|\, v]$ generates a $[q+2,3]$ MDS code, which has the only solutions $v= a(0,1,0)^T, a \neq 0$  as noted above.
\end{enumerate}

\subsection{Remarks on the Covering Radius of MDS codes} \label{sec_cov_MDS}
It is sometimes asserted in literature (for example \cite{BGP}, \cite{GK98}) that the covering radius of any linear $[n,k]$ MDS code $C$  is either $n-k$ or $n-k-1$. In this section we wish to emphasize that there is no known correct proof of this assertion, and hence  it remains  a widely believed conjecture. Let $A$ and $A^{\perp}$ denote a pair of generator and parity check matrices for $C$. We recall that the covering radius of  any linear $[n,k]$ code is at most $n-k$. 
\begin{lem} \label{lem_supercode} The following assertions are equivalent for an $[n,k]$ MDS code $C$:
\begin{enumerate}
\item $\rho(C)=n-k$
\item There exists a word $u \in \f{q}^n$ such that the $(k+1) \times n$ matrix $(\bbsm A \\  u \besm)$ generates a $[n,k+1]$ MDS code.
\item There exists a word $u \in \f{q}^n$ such that the $(k+1) \times (n+1)$ matrix $(\bbsm A & 0 \\ u & 1 \besm)$ generates a $[n+1,k+1]$ MDS code.
\item There exists a vector $v \in \f{q}^{n-k}$ such that the $(n-k) \times (n+1)$ matrix 
$(A^{\perp} \,|\, v)$ generates a $[n+1,n-k]$ MDS code extending $C^{\perp}$ by one coordinate.
\end{enumerate}   
\end{lem}
\begin{IEEEproof} $(1) \Leftrightarrow (2)$: We have $\rho(C)=n-k$ if and only if there exists a  word $u \in \f{q}^n$  with error distance $n-k$. This in turn is true if and only if no $(k+1) \times (k+1)$ minor of the matrix  $(\bbsm A  \\ u \besm)$ is zero.\\
$(2) \Leftrightarrow (3)$:  Since every $k \times k$ minor of $A$ is nonzero, it follows that every $(k+1) \times (k+1)$ minor of the matrix  $(\bbsm A & 0 \\ u & 1 \besm)$ is nonzero if and only if the same is true of the matrix  $(\bbsm A \\  u \besm)$.\\
$(3) \Leftrightarrow (4)$:  Suppose $(\bbsm A & 0 \\ u & 1 \besm)$  generates a $[n+1,k+1]$ MDS code. A parity check matrix for this code is $(A^{\perp} \,|\,  {-A}^{\perp}u^t )$. Since the dual code of an MDS code is MDS, it follows that  the 
latter matrix generates an $[n+1,n-k]$ MDS code.  Conversely suppose  the $(n-k) \times (n+1)$ matrix $(A^{\perp} \,|\,  v )$ generates an MDS code. Since $A^{\perp}$ is full rank, there exists a word $u \in \f{q}^n$ such that  $A^{\perp} u^t=-v$. The
$(k+1) \times (n+1)$ matrix $(\bbsm A & 0 \\ u & 1 \besm)$ also generates an MDS code as it is  a parity check matrix for the code generated by $(A^{\perp} \,|\,  v )$.
\end{IEEEproof} \medskip

If the $(k+1) \times n$ matrix $(\bbsm A \\ u \besm)$ generates a $[n,k+1]$ code, then the latter code is called a supercode containing $C$. By Lemma \ref{lem_supercode} it follows that $\rho(C) < n-k$  if and only if $C$ cannot be embedded in an MDS supercode, or equivalently if $C^{\perp}$ cannot be extended to a $[n+1,n-k]$ MDS code. It is not true in general that any $[n,k]$ MDS code with $n \leq q$ can be embedded in a $[n,k+1]$ MDS supercode, because dually, it is not true that any $[n,n-k]$ MDS code for $n \leq q$ can be extended to a $[n+1,n-k]$ MDS code. In finite geometry terms there do exist \emph{complete} arcs of length $n \leq q$ in $\bP^{n-k-1}(\f{q})$ for some $k,n$. For instance  several examples of complete $n$-arcs in $\bP^2(\f{q})$ are known for some $q,n$ with $n \leq q$ (see \cite[Tables 1,2]{Hir}), which means that there do exist $[n,n-3]$ MDS codes for $n<q$ which cannot be embedded in a MDS supercode. Thus, Remark 1) of \cite{GK98} is not accurate. 
For such codes $C$, it is unlikely that $\rho(C) < n-k-1$, however there is no known proof that $\rho(C)= n-k-1$ to the authors' best knowledge.\\

Theorem 2 of \cite{GK98} asserts that any $[q+1,k]$ MDS code  $C$ (except the cases  $k \in \{2,q-2\}$ when $q$ is even) has covering radius $q-k$. Here, by Lemma \ref{lem_supercode},  one must clearly add the hypothesis that $C$ cannot be embedded in a MDS supercode, or equivalently $C^{\perp}$ cannot be extended to an MDS code of length $q+2$ (the additional hypothesis is not necessary if
 the MDS conjecture is true in dimension $q+1-k$).  In the proof of this theorem, the authors assume that $C$ can be taken to be  $PRS(k)$, for which the result is true by Lemma \ref{lem_cover}. However, it is not true in general that every MDS code of length $q+1$ is PRS, and therefore we cannot conclude that $\rho(C)=q-k$. It is unlikely that $\rho(C) <q-k$ but there is no proof yet.


\subsection{Automorphisms of PRS codes} \label{sec_aut}
The automorphism group of a linear code acts on the set of deep holes of the code, and hence can be a useful tool to determine the set of deep holes. We begin with some general notions concerning automorphisms of a linear code. We recall that  the subgroup of $GL_n(\f{q})$ consisting of Hamming isometries of $\f{q}^n$ is the group of $n \times n$ monomial matrices (a $n \times n$ monomial matrix is a product of a $n \times n$ permutation matrix and a $n \times n$ diagonal matrix). Since we are writing words of $\f{q}^n$ as row vectors, the action of $A \in GL_n(\f{q})$ on a word $u$ is  $u \mapsto u A^{-1}$.  For a linear $[n,k]$ code $C$, the automorphism group Aut$(C)$ of the code is the subgroup of  $GL_n(\f{q})$ consisting of those monomial matrices $A$ satisfying $cA^{-1} \in C$ for all $c \in C$.  Since $C$ is linear, the group of scalar matrices $\{\lambda I_n : \lambda \in \f{q}^{\times}\}$ (where $I_n$ is the $n\times n$ identity matrix) is contained in the center of Aut$(C)$. Let $\bP\text{Aut}(C)$ denote the quotient group $\text{Aut}(C) /\{\lambda I_n : \lambda \in \f{q}^{\times}\}$.\\

Given a pair $G, H$ of generator and parity check matrices for $C$, we can define monomorphisms $\imath:\text{Aut}(C) \hookrightarrow GL_k(\f{q})$ and $\jmath:
\text{Aut}(C) \hookrightarrow GL_{n-k}(\f{q})$  defined as follows. For each $A \in \text{Aut}(C)$  the matrix $GA^{-1}$ is also a generator matrix for $C$. The fact that $A^{-1}$ also is in $\text{Aut}(C)$ implies that $GA$ is also a generator matrix for $C$. Similarly,  the fact that $c A^{-1} \in C$ implies that  $HA^{-t}c^t=0$ for all $c \in C$. Therefore,   $HA^{-t}$ is also a parity check matrix for $C$. Since generator and parity check  matrices are unique upto row equivalence, it follows that there exist  matrices $A' \in GL_k(\f{q})$ and $A'' \in GL_{n-k}(\f{q})$ such that $A'G=GA$ and $A''H = H A^{-t}$. Moreover, the matrices $A'$ and  $A''$ are unique because $G$ and  $H$ are full rank. We define 
\[ \imath(A)=A' \quad \text{ and} \quad  \jmath(A)=A''.\] The identities 
\[ GAB=\imath(A)GB=\imath(A)\imath(B)G,\qquad   H A^{-t} B^{-t}=\jmath(A) HB^{-t}=\jmath(A)\jmath(B) H,\] show that $\imath,\jmath$  are group homomorphisms. Again, the fact that $G$ and  $H$ are full rank implies that the only matrices $A', A''$ satisfying $A'G=G$ and $A''H=H$ are the identity matrices. Therefore, 
$\imath$ and $\jmath$  are  monomorphisms. Finally, we use the same notation $\imath$ and  $\jmath$ for  the induced monomorphisms $\imath: \bP\text{Aut}(C) \hookrightarrow PGL_k(\f{q})$ and $\jmath: \bP\text{Aut}(C) \hookrightarrow PGL_{n-k}(\f{q})$. Here $PGL_m(\f{q})$ denotes, as above  the quotient group $GL_m(\f{q})/\{\lambda I_m : \lambda \in \f{q}^{\times}\}$.  Since $A$ carries $C$ to $C$,  it follows that Aut$(C)$  acts on the vector space $\f{q}^n/C$ of cosets of $C$.  The action of Aut$(C)$  on  $\f{q}^n/C$,  induces an action of $\bP\text{Aut}(C)$ on the projective space $\bP(\f{q}^n/C)$ of equivalence classes of non-codewords. The bijective correspondence $\text{syn}:\bP(\f{q}^n/C) \to \bP(\f{q}^{n-k})$ (given in  \eqref{eq:projsyn})  respects the  action of $\bP\text{Aut}(C)$:
\begin{equation} \label{eq:aut_syn}  \text{syn}(u A^{-1})=H A^{-t} u^t=\jmath(A) Hu^t=\jmath(A) \text{syn}(u). \end{equation}
For $A \in \text{Aut}(C)$ and a received word $u$, clearly
the error distance   $d(u,C)=d(uA^{-1}, C)$. In particular $\text{Aut}(C)$ acts on the set of deep holes of $C$, and 
 $\bP\text{Aut}(C)$ acts on the set of deep hole classes of $C$. \\

We now return to the code $C=PRS(k)$.  For $2 \leq k \leq q-2$, if $A \in \text{Aut}(PRS(k))$, then the equation $\imath(A) G = G A$ implies that $\imath(A) \in PGL_k(\f{q})$ preserves the set of $(q+1)$ points $\{c_k(t) : t \in \f{q} \cup \infty\} \subset \bP^{k-1}(\f{q})$. Similarly, the equation $\jmath(A) H = H A^{-t}$  implies that $\jmath(A) \in PGL_{q+1-k}(\f{q})$  preserves the set of $(q+1)$ points
$\{c_{q+1-k}(t) : t \in \f{q} \cup \infty\} \subset \bP^{q-k}(\f{q})$. Here we have used the fact that for a monomial matrix $A$, the columns of a matrix $MA$ are obtained from the columns of $M$ by permutation and rescaling.  We recall that there is a bijection $\f{q} \cup \infty \to \bP^1(\f{q})$ given by $t \mapsto (1:t)$ where it is understood that  $(1:t)=(0:1)$ for $t =\infty$. The action of $GL_2(\f{q})$ on the vector space $\f{q}^2$ induces an action of $PGL_2(\f{q})$ on the set $\bP^1(\f{q})$ of one dimensional subspaces of $\f{q}^2$.   Given  $g= (\bbsm a & b \\ c & d \besm) \in PGL_2(\f{q})$ and $(1:t) \in \bP^1(\f{q})$, we have $ (\bbsm a & b \\ c & d \besm) (\bbsm 1\\t \besm)=(\bbsm a+bt \\ c+dt \besm)$. In terms of the identification 
$\f{q} \cup \infty \to \bP^1(\f{q})$, this is usually written as $g(t)=(c+dt)/(a+bt)$ and referred to as a  M\"{o}bius  or fractional linear transformation.
\begin{definition} \label{def_gm}
For each $2 \leq m \leq q$,  we define functions
$GL_2(\f{q}) \to  GL_m(\f{q})$ denoted $g \mapsto g_m$  as follows. For $g = (\bbsm a & b \\ c & d \besm)$,  the 
$ij$-th entry of $g_m$  is the coefficient of $X^{j-1}$ in the polynomial $(a+b X)^{m-i} (c+d X)^{i-1}$. 
\end{definition}
 For example  $g_2=g$ and,
 \begin{IEEEeqnarray}{rLlr}
 \nonumber g_3=&\begin{pmatrix} a^2 & 2ab & b^2\\ ac & ad+bc &  bc \\c^2 & 2cd & d^2   \end{pmatrix}
  &\text{ for $g=\begin{pmatrix} a & b \\ c & d \end{pmatrix}$,}\\ 
 \label{eq:reciprocal} g_m=& \begin{pmatrix}   & & &1 \\ &&1& \\ &\iddots&& \\
1&&&  \end{pmatrix}  &\text{ for $g=\begin{pmatrix} 0 & 1 \\ 1 & 0 \end{pmatrix}$,}\\
\label{eq:translate} g_m=& \begin{pmatrix} 1 & & & \\ c &1&&& \\ c^2&2c&1&&\\ 
\vdots&\vdots&\vdots&\ddots& \\
c^{m-1}& (m-1) c^{m-2}&\textstyle\binom{m-1}{2} c^{m-3}&\hdots&1  \end{pmatrix}  &\text{ for $g=\begin{pmatrix} 1 & 0 \\ c & 1  \end{pmatrix}$}.\end{IEEEeqnarray}

 We collect some properties of the matrices $g_m$: 
 \begin{enumerate}
\item  \cite[Proposition 2.6]{BGHK17}: the map $g \mapsto g_m$ is a group homomorphism  and the induced homomorphism $PGL_2(\f{q}) \to PGL_m(\f{q})$ (which we again denote by $g \mapsto g_m$) is a monomorphism.
\item  \cite[Proposition 2.5]{BGHK17}):  For each $t \in \f{q} \cup \infty$ we have
\begin{equation} \label{eq:aut_rnc} g_m c_m(t) = c_m(g(t)) \in \bP^{m-1}(\f{q}) \end{equation} 
\item \cite[Theorem 2.10]{BGHK17}:  For $m < q$, the only elements of $PGL_m(\f{q})$ which preserve the set  
$\{c_m(t) : t \in \f{q} \cup \infty\} \subset \bP^{m-1}(\f{q})$
are  $\{g_m : g \in PGL_2(\f{q})\}$   
\end{enumerate}
Thus for $C=PRS(k)$, the images of the  monomorphisms $\imath:\bP\text{Aut}(C) \hookrightarrow PGL_k(\f{q})$ and $\jmath:\bP\text{Aut}(C) \hookrightarrow PGL_{n-k}(\f{q})$ are precisely $\{g_k : g \in PGL_2(\f{q})\}$ and $\{g_{q+1-k} : g \in PGL_2(\f{q})\}$.
The group  $\bP\text{Aut}(C)$ itself can be described as follows:
The action of $PGL_2(\f{q})$ on $\f{q} \cup \infty$  gives a monomorphism $g \mapsto \Pi(g)$ from $PGL_2(\f{q})$ to the group of permutation matrices in $GL_{q+1}(\f{q})$ defined by: 
\[ [g \alpha_1, \dots, g \alpha_{q+1}]=[ \alpha_1, \dots,  \alpha_{q+1}] \Pi(g).\]
By the identity \eqref{eq:aut_rnc}, it follows that  there exists a  diagonal matrix $\Delta_m(g)$  such that the $n \times n$ monomial matrix  $B_m(g)= \Pi(g) \Delta_m(g)$ satisfies the property 
\[  g_m G_m=G_m B_m(g).\] In particular, 
\[ \imath(B_k(g))=g_k, \quad \text{ and}\quad  \jmath(B_k(g))=g_{q+1-k}.\]
 Since  $g \mapsto g_k$ (from $PGL_2(\f{q}) \to PGL_k(\f{q})$)  and $\imath:\bP\text{Aut}(PRS(k)) \to PGL_k(\f{q})$ are both monomorphisms, it follows that $g \mapsto B_k(g)$ is an isomorphism from $PGL_2(\f{q})$ to $\bP\text{Aut}(PRS(k)) \subset PGL_{q+1}(\f{q})$.\\

For completenes, we write down the matrices $\Delta_m(g)=\text{diag}(\delta_1, \delta_2, \cdots, \delta_{q+1})$:
\begin{align} \label{eq:def_delta}
\delta_i=\begin{cases}
(a+b\alpha_i)^{m-1} &\mbox { if $\alpha_i\neq -\frac{a}{b}, \infty$},\\
(c-d\frac{a}{b})^{m-1} &\mbox{ if  $b \neq 0$, $\alpha_i= -\frac{a}{b}$},\\
b^{m-1}  &\mbox{ if $b \neq 0$, $\alpha_i= \infty$},\\
d^{m-1}  &\mbox{ if $b=0$, $\alpha_i= \infty$}.\\
\end{cases}
\end{align} 
Since $G_k$ is also a parity check matrix for $PRS(q+1-k)$, it follows from the definition of the homomorphism $\jmath$ that
\[ G_k {B_{q+1-k}(g)}^{-t}= \jmath(B_{q+1-k}(g)) G_k = g_k G_k = G_k B_k(g).\]
Therefore,
 \[ B_{q+1-k}(g)= B_k(g)^{-t}. \]
Using this, the equation  \eqref{eq:aut_syn} for $C=PRS(k)$ becomes:
\[   \text{syn}(u B_k(g^{-1}))=G_{q+1-k} B_{q+1-k}(g) u^t=g_{q+1-k}  G_{q+1-k} u^t= g_{q+1-k} \text{syn}(u).
\]
We summarize this in the following lemma:
\begin{lem} \label{lem:aut_syn_PRS}
Let $u$ and $v$ be deep hole classes of $PRS(k)$. Then $v$ is in the $\bP \text{Aut}(PRS(k))$ orbit of $u$ if and only if there exists $g \in PGL_2(\f{q})$ such that 
$g_{q+1-k} \mathrm{syn}(u) = \mathrm{syn}(v)$. 
\end{lem}
We end this section with a calculation of the $PGL_2(\f{q})$ orbit of 
 \begin{equation} \label{eq:N_def} 
N_m= (0:\cdots:0:1:0) \in \mathbb{P}^{m-1}(\f{q}), \quad m \geq 3,
 \end{equation}
which we need in the next section. We also use the same symbol $N_m$ for the vector $(0,\dots,0,1,0)^T \in \f{q}^m$.
  \begin{lem}\label{stabofN}
Let  $3 \leq m \leq q$, and let $N_m \in \mathbb{P}^{m-1}(\f{q})$ be as above.
\begin{enumerate}
		\item if $m=3$ and $q$ is odd, the orbit of $N_m$ has size $q(q+1)/2$ and its stabilizer is the group $\{t \mapsto d t^{\pm 1} : d \in \f{q}^{\times}\}$.
		\item if  $m>3$ and $m\not\equiv 1\bmod p$, the orbit of $N_m$ has size $q(q+1)$ and its stabilizer is the group $\{t \mapsto dt : d \in \f{q}^{\times}\}$. 
	\item if  $m>3$ and $m \equiv 1\bmod p$, the orbit of $N_m$ has size $(q+1)$ and its stabilizer is the group
 $\{t \mapsto dt+c : d \in \f{q}^{\times}, c \in \f{q}\}$.
	\item if $m=3$ and $q$ is even, the orbit of $N_m$ has size $1$ and its stabilizer is the whole group $PGL_2(\f{q})$.
	\item $N_m + c_m(\infty)$ is in the $PGL_2(\f{q})$-orbit of $N_m$ if and only if $m\not\equiv 1\bmod p$. In case $m\equiv 1\bmod p$, the orbit of $N_m + c_m(\infty)$ has size $q^2-1$, and its stabilizer  is the group $\{t \mapsto t+c :  c \in \f{q}\}$.
\end{enumerate}
\end{lem}
\begin{IEEEproof}
 For  $g=(\bbsm a & b\\ c& d \besm) \in PGL_2(\f{q})$, we  have  by Definition \ref{def_gm}:
	\begin{equation} \label{eq:orbN_k}
	g_m \cdot N_m = \left( \bbsm
	(m-1)ab^{m-2}\\cb^{m-2}+(m-2)ab^{m-3}d\\2cb^{m-3}d+(m-3)ab^{m-4}d^2\\\vdots \\ (m-2)cbd^{m-3}+ad^{m-2}\\ (m-1)cd^{m-2} 	\besm\right).
	\end{equation}
		In order to determine when  this  equals $N_m$, we consider the cases $m \equiv 1 \bmod p$ and $m\not\equiv 1\bmod p$  separately. First suppose  $m\not\equiv 1\bmod p$.  The first and last components of \eqref{eq:orbN_k} imply $ab=cd=0$, i.e. either $a=d=0$ or $b=c=0$. In the former case $g_m N_m=(0:1:0:\cdots:0)$ which equals $N_m$ if and only if $m=3$. If $b=c=0$, then 	$g_m N_m=N_m$. This proves the assertions 1) and 2).   Now suppose   $m\equiv 1\bmod p$.  If $b=0$, we have  $g_m N_m =N_m$.  If $b \neq 0$, using the fact that $ad-bc \neq 0$, we can write 
		 \[ g_m N_m= (0:1:2d/b:3 (d/b)^2: \dots: (m-2) (d/b)^{m-3}: 0).\] This equals $N_m$ if and only if $m=3$. This proves the assertions 3) and 4).\\

If  $m\not\equiv 1 \bmod p$, then $g_m N_m = N_m + c_m(\infty)$ for $g(t)=t+(m-1)^{-1}$. If  $m \equiv 1 \bmod p$, then it is clear from \eqref{eq:orbN_k} that, for every $v$ in the $PGL_2(\f{q})$-orbit of $N_{m}$, the last entry of $v$ is zero. In particular, $N_m + c_m(\infty)$ is not in the $PGL_2(\f{q})$-orbit of $N_{m}$. 
Also, for $m \equiv 1 \bmod p$, we have 
\[  g_m (N_m + c_m(\infty)) =\left( \bbsm
	b^{m-1}\\  b^{m-2}d\\ b^{m-3}d^2\\ \vdots\\  b d^{m-2}\\ d^{m-1}	\besm  \right)
	+ (bc-ad) \left( \bbsm 0 \\ b^{m-3}\\2b^{m-4}d\\\vdots\\(m-2)d^{m-3}  \\ 0\besm 	\right).\]
Thus $g$ stabilizes $N_m + c_m(\infty)$ if and only if $b=0$ and $a=d$. Therefore, the  stabilizer of $N_m + c_m(\infty)$ is the group $\{t \mapsto t+c :  c \in \f{q}\}$.  This proves assertion 5).
\end{IEEEproof}

\subsection{New deep holes of PRS codes} \label{sec_new_deepholes}
In this section we obtain the two new deep hole classes of $PRS(k)$ given in Theorem \ref{thm:Rotationsofdegk} and 
Theorem \ref{thm:main_irr2}. We throughout assume $2 \leq k \leq q-3$  in this section.
\subsubsection*{Proof of Theorem \ref{thm:Rotationsofdegk}}
 Assuming $\rho(PRS(k))=q-k$, we need to show that  the  $q^2$ words  
\[u(i,a)=
	(\tfrac{1}{\alpha_1-\alpha_i},\cdots,\tfrac{1}{\alpha_{i-1}-\alpha_i},a,\tfrac{1}{\alpha_{i+1}-\alpha_i},\cdots, \tfrac{1}{\alpha_q-\alpha_i}, 0), \quad  1 \leq i \leq q, \; a \in \f{q}	\]
represent distinct deep holes classes of $PRS(k)$, and that these are distinct from the $q$ deep holes  of Theorem \ref{deephole:degreek}. 
 The $j$-th component of $\text{syn}(u(i,a))=G_{q+1-k} u(i,a)^T$ is 
\[  a \alpha_i^{j-1} +\sum_{\ell \neq i} \tfrac{\alpha_{\ell}^{j-1}}{\alpha_{\ell} - \alpha_i}. \]
Expanding $\alpha_{\ell}^{j-1}$ as $(\alpha_{\ell} - \alpha_i + \alpha_i)^{j-1}$, we have:
 \[  \tfrac{\alpha_{\ell}^{j-1}}{\alpha_{\ell} - \alpha_i} -  (j-1) \alpha_i^{j-2} = \sum_{s \neq 1} \textstyle\binom{j-1}{s} \alpha_i^{j-1-s} \, (\alpha_{\ell}- \alpha_i)^{s-1}. \]
Summing the last equation over all $\ell \neq i$, we get:
\[ (j-1) \alpha_i^{j-2}+  \sum_{\ell \neq i} \tfrac{\alpha_{\ell}^{j-1}}{\alpha_{\ell} - \alpha_i} =0,\]
where we have used the identity \eqref{eq:sumofpowers}, and the fact that $s-1 <j \leq q+1-k \leq q-1$. Therefore, the $j$-th component of $\text{syn}(u(i,a))$ is
\[  a \alpha_i^{j-1} -(j-1) \alpha_i^{j-2}.\]
 In other words:
 \begin{equation} \label{eq:syn_u_i_a}  \text{syn}(u(i,a)) = a \,c_{q+1-k}(\alpha_i)-  c_{q+1-k}'(\alpha_i)  \end{equation}
where $c_{q+1-k}'(t)=(0,1,2t,3t^2, \dots, (q-k) t^{q-k-1})^T$.\\

For $g=(\bbsm 1 &0\\c &1 \besm) \in GL_2(\f{q})$,  it follows from 
\eqref{eq:translate} that:
\begin{equation} \label{eq:def_delta1}
g_m  c_m(X)= c_m(X+c),
\end{equation}
where each of the $m$ components of this equation are polynomial identities in $\f{q}[X]$ with  $c_m(X)=(1,X,X^2, \dots,X^{m-1})^T$. Differentiating this polynomial identity with respect to $X$ gives 
\[g_m c_m'(X)=c_m'(X+c), \quad \text{ where }  \, c_m'(X)=(0,1,2X, \dots,(m-1)X^{m-2})^T.\]
 Using this in \eqref{eq:syn_u_i_a}, we get 
\begin{equation} \label{eq:syn_translate} g_{q+1-k} \text{syn}(u(i,a))= a c_{q+1-k}(0)-  c_{q+1-k}'(0)=(a,-1,0\dots,0)^T \quad \text{for $g=(\bbsm 1 &0\\ -\alpha_i &1 \besm)$}.\end{equation}
Further, using \eqref{eq:reciprocal} we get:
\[ h_{q+1-k} \text{syn}(u(i,a))=(0,\dots,0,-1,a)^T= -N_{q+1-k} + a \,c_{q+1-k}(\infty) \quad \text{for $h=(\bbsm 0 &1 \\1 &0 \besm) 
(\bbsm 1 &0\\ -\alpha_i &1 \besm)$}.\]
Thus the projective syndrome $\text{syn}(u(i,a))$ is in the $PGL_2(\f{q})$-orbit of 
$N_{q+1-k}-a c_{q+1-k}(\infty)$.  Since $N_{q+1-k}-a c_{q+1-k}(\infty)$ is the syndrome of the deep hole $(\alpha_1^k, \alpha_2^k, \dots, \alpha_q^k,a)$, it follows from Lemma \ref{lem:aut_syn_PRS} that $u(i,a)$ are deep holes. \\

Next we show that the $q^2$ words $\{u(i,a) : 1 \leq i \leq q, a \in\f{q}\}$ represent distinct deep hole classes. Suppose the projective syndromes $\text{syn}(u(i,a))=\text{syn}(u(j,b))$. In view of \eqref{eq:syn_translate}, we may assume $\text{syn}(u(j,b))=(b:-1:0:\dots:0)$ (i.e. $\alpha_j=0$). If $\alpha_i=0$, then the expression
$\text{syn}(u(i,a))=(a:-1:0:\dots:0)$  shows that $b=a$, and hence $u(i,a)=u(i,b)$.
Next, suppose  $\alpha_i \neq 0$.  Since  $q+1-k \geq 4$, the last two components of $(b:-1:0:\dots:0)$  are zero, but the last two components of $\text{syn}(u(i,a))$, namely 
\[ \alpha_i^{q-k-2} \left(a \alpha_i - (q-k-1) \right), \quad \alpha_i^{q-k-1} \left( a \alpha_i - (q-k) \right),\] cannot both be zero.  This contradiction shows that the projective syndromes $\text{syn}(u(i,a))$ and $ \text{syn}(u(j,b))$ are distinct if  $i \neq j$.\\

Next we show that the deep hole classes represented by $u(i,a)$ are distinct from the the $q$ classes of Theorem \ref{deephole:degreek}. The fact that  $q+1-k \geq 4$ implies that the first two components of $N_{q+1-k}- b \,c_{q+1-k}(\infty)=(0:\dots:0:1:-b)$ are zero, but the first two components
of $\text{syn}(u(i,a)) = (a: a \alpha_i-1: \dots)$ cannot both be zero. This shows that the deep hole classes of the words $u(i,a)$ are distinct from the $q$ classes of Theorem \ref{deephole:degreek}. \\

We recall  from  Lemma \ref{stabofN}, that $\{N_{q+1-k}-a c_{q+1-k}(\infty) : a \in \f{q}\}$  is in the $PGL_2(\f{q})$ orbit of $N_{q+1-k}$ provided   $p \nmid k$.
If $p \mid k$, then the $PGL_2(\f{q})$ orbits of $N_{q+1-k}$  and $N_{q+1-k}+c_{q+1-k}(\infty)$ are distinct, and the latter orbit contains
$\{N_{q+1-k}-a c_{q+1-k}(\infty) : a \in \f{q}^{\times}\}$. Combining this with the fact that $\text{syn}(u(i,a))$ is in the $PGL_2(\f{q})$ orbit of $N_{q+1-k}-a c_{q+1-k}(\infty)$, we conclude:\\
The $q^2+q$ deep holes of Theorems \ref{deephole:degreek} and Theorem \ref{thm:Rotationsofdegk} put together form: \begin{enumerate}
\item in case  $p \nmid k$,  the $PGL_2(\f{q})$ orbit  of $(\alpha_1^k, \alpha_2^k, \dots, \alpha_q^k,0)$
\item   in case  $p \mid k$,  the $PGL_2(\f{q})$ orbits of $(\alpha_1^k, \alpha_2^k, \dots, \alpha_q^k,0)$ and $(\alpha_1^k, \alpha_2^k, \dots, \alpha_q^k,1)$ of sizes  $q+1$ and $q^2-1$ respectively.  \\
\end{enumerate}
\emph{Remark}: In terms of the the standard degree $(q-k)$  normal rational curve in $\bP^{q-k}(\f{q})$, the projective tangent line to the curve at $c_{q+1-k}(t)$ for each $t \in \f{q} \cup \infty$ has $(q+1)$ points with $\f{q}$-coordinates given by $c_k(t)$ itself and the $q$ points $\{c_k'(t) -a c_k(t) : a \in \f{q}\}$. If $t=\infty$, then these  $(q+1)$ points are $c_{q+1-k}(\infty)$  and $\{N_{q+1-k}- a c_{q+1-k}(\infty) : a \in \f{q}\}$.  As shown above, 
the tangent lines have  no pairwise intersection when   $k \leq q-3$. Thus the \emph{geometric interpretation} of the $(q+1)q$ syndromes of the  deep hole classes in Theorems \ref{deephole:degreek}  and   \ref{thm:Rotationsofdegk}, is that these consist of those points with $\f{q}$-coordinates which are not on the curve, but are  in the union of the tangent lines to the curve. $\qquad \blacksquare$\\

\subsubsection*{Proof of Theorem \ref{thm:main_irr2}}
For each  $p(X)$ in the set of $(q^2-q)/2$ monic irreducible quadratic polynomials over $\f{q}$, and for each $a \in \f{q} \cup \infty$, let $u(a,p(X))$ be the word in $\f{q}^{q+1}$ defined by 
\begin{equation} \label{eq:u_a_P}
u(a,p(X))=\begin{cases} 
(\frac{1}{p(\alpha_1)}, \dots, \frac{1}{p(\alpha_q)},0)  &\text{ if $a=\infty$} \\
 (\frac{\alpha_1+a}{p(\alpha_1)},\frac{\alpha_2+a}{p(\alpha_2)},\cdots,\frac{\alpha_q+a}{p(\alpha_q)},0)  &\text{ if  $a \in \f{q}$} 
\end{cases}
\end{equation}
 Assuming $\rho(PRS(k))=q-k$, we must show that $(q+1)q(q-1)/2$ words given by 
$u(a,p(X))$ represent distinct deep hole classes of $PRS(k)$ for  $2\leq k\leq q-3$, and that these are distinct from the $q^2+q$ classes of Theorems \ref{deephole:degreek}  and \ref{thm:Rotationsofdegk}. We begin with two lemmas.
\begin{lem}\label{synoffrac}
The projective syndrome of the 	word $u(a,p(X))$ is
	\[
	\mathrm{syn}(u(a,p(X)))=\mu_a \, c_{q+1-k}(\mu)+\mu_a^q \, c_{q+1-k}(\mu^q),
	\]
	where $\mu$ is a root of $p(X)$  in a quadratic extension $\f{q^2}$ of $\f{q}$, and 
	\begin{equation} \label{eq:lambda_a}
	 \mu_a= \begin{cases}  \mu +a  &\text{ if $a \in \f{q}$}\\
                                            1  &\text{ if $a=\infty$} \end{cases}    \end{equation}
\end{lem}
\begin{IEEEproof}
Let $\mu\in\f{q^2}$ be a root of $p(X)$, and let $\sigma$ denote the nontrivial automorphism $x \mapsto x^q$ of $\f{q^2}$ over $\f{q}$. We have:
\begin{IEEEeqnarray}{lCr} \label{eq:partial_frac}
	\frac{1}{p(X)} &=\frac{1}{\mu-\mu^q} \frac{1}{X-\mu}+
	\sigma ( \frac{1}{\mu-\mu^q} \frac{1}{X-\mu}), \\ \nonumber
	\frac{X+a}{p(X)}&=\frac{\mu+a}{\mu-\mu^q} \frac{1}{X-\mu}+\sigma (\frac{\mu+a}{\mu-\mu^q} \frac{1}{X-\mu}).
	\end{IEEEeqnarray}
Using this in \eqref{eq:u_a_P}, we get:
\[ \text{syn}(u(a,p(X)) )=G_{q+1-k} u(a,p(X))^T= \begin{cases} \frac{1}{\mu-\mu^q} w + \sigma(\frac{1}{\mu-\mu^q} w) 
 &\text{ if $a=\infty$} \\
 \frac{\mu+a}{\mu-\mu^q} w + \sigma(\frac{\mu+a}{\mu-\mu^q} w)  &\text{ if $a \in \f{q}$}
 \end{cases},\]
where 
\[ w=G_{q+1-k} \left(\begin{smallmatrix}
	\tfrac{1}{ \alpha_1-\mu}\\ \vdots\\ \tfrac{1}{\alpha_q-\mu} \\ 0 	\end{smallmatrix}\right)= 
\sum_{\alpha\in\f{q}}	\left(\begin{smallmatrix}
	 \tfrac{1}{ \alpha-\mu}\\  \tfrac{\alpha}{\alpha-\mu}\\ \vdots\\ \tfrac{\alpha^{q-k}}{\alpha-\mu}	\end{smallmatrix}\right).\]
Using the partial fraction expansions
\[ \frac{X^j}{X^q-X}= \sum_{\alpha \in \f{q}} \frac{-\alpha^j}{X-\alpha},\] 
we get
\[w= 
	\frac{1}{\mu^q-\mu} \left(\begin{smallmatrix}
	1\\ \mu \\ \vdots\\ \mu^{q-k}	\end{smallmatrix}\right) =  	\frac{1}{\mu^q-\mu}  c_{q+1-k}(\mu).
	\]

Multiplying $\text{syn}(u(a,p(X)) )$ by $-(\mu-\mu^q)^2 \in \f{q}^{\times}$ does not change the projective syndrome. Therefore,
\[ \text{syn}(u(a,p(X)) )=\mu_a \, c_{q+1-k}(\mu) + \sigma (\mu_a \, c_{q+1-k}(\mu))  \in \bP^{q-k}(\f{q}), \]
where $\mu_a$ is as defined  in \eqref{eq:lambda_a}. \end{IEEEproof}
\begin{lem}\label{aut_synoffrac}
The group $\bP \text{Aut}(PRS(k))$ preserves the set of $(q+1)q(q-1)/2$ words of the form $u(a,p(X))$ .
\end{lem}
\begin{IEEEproof}
We know from \eqref{eq:def_delta} that 
\[g_{q+1-k} (\mu_a c_{q+1-k}(\mu) + \sigma (\mu_a c_{q+1-k}(\mu)) = \lambda  c_{q+1-k}(\nu)+\lambda^q c_{q+1-k}(\nu^q),\]
where  $\nu=(\gamma+ \delta \mu)/(\alpha + \beta \mu)$, and  $\lambda  = (\alpha + \beta \mu)^{q-k} \mu_a$.
Since $\mu\in \f{q^2} \setminus \f{q}$, the same is true for $\nu$, and hence 
any element of $\f{q^2}$ is of the form $ r + s \nu$ for some $r,s \in \f{q}$. In particular any element of $\f{q^2}^{\times}/\f{q}^{\times}$ is represented by one of the $(q+1)$ elements 
\[\nu_b= \begin{cases}  \nu +b  &\text{ if $b \in \f{q}$}\\
                                            1  &\text{ if $b=\infty$} \end{cases}. \]
Thus we may take $\lambda=\nu_b$ for some $b \in \f{q} \cup \infty$.
\end{IEEEproof} \medskip

Next we show that the $u(a,p(X))$ are deep holes of $PRS(k)$ when $\rho(PRS(k))=q-k$. By Lemma \ref{coveringradiusfromparitycheck}, we must show that $\text{syn}( u(a,p(X)))$ is not in the $\f{q}$-span of $(q-1-k)$ columns of $G_{q+1-k}(\f{q})$.  
Consider the  $(q+1-k) \times (q^2+1)$ matrix  \[ G_{q+1-k}(\f{q^2})=[c_{q+1-k}(t_1)| \dots | c_{q+1-k}(t_{q^{2}+1})],\]  where $t_1, \dots, t_{q^2+1}$ is a listing of $\f{q^2} \cup \infty$.  We know that  any $(q+1-k)$ columns of this matrix are linearly independent over $\f{q^2}$. In particular 
\[ \text{syn}(u(a,p(X)))=\mu_a c_{q+1-k}(\mu)+\mu_a^q c_{q+1-k}(\mu^q), \]
 is not in the $\f{q}$-span of $(q-1-k)$ columns of $G_{q+1-k}(\f{q})$, as was to be shown. 
Moreover  for $k  \leq q-3$, we have $4 \leq q-k+1$, and hence any  four columns of $G_{q+1-k}(\f{q^2})$ are linearly independent over $\f{q}$. This shows that if $(a,p(X)) \neq (b,\tilde p(X))$ then their projective syndromes
\[ \mu_a c_{q+1-k}(\mu)+\mu_a^q c_{q+1-k}(\mu^q) \quad \text{and} \quad \nu_{b} c_{q+1-k}(\nu)+ \nu_{b}^q c_{q+1-k}(\nu^q), \]
 (where $\tilde p(X)=(X- \nu)(X- \nu^q)$) are distinct. This proves that the  $(q+1)q(q-1)/2$ words $u(a,p(X))$ of  \eqref{eq:u_a_P} represent distinct deep hole 
classes of $PRS(k)$. \\

Next, we show that the $q^2+q$ deep hole classes of Theorems \ref{deephole:degreek} and \ref{thm:Rotationsofdegk} are distinct from 
those of   Theorem \ref{thm:main_irr2}.  The projective syndromes of the former deep hole classes are in the $PGL_2(\f{q})$-orbit of $N_{q+1-k}$ or $N_{q+1-k}+ c_{q+1-k}(\infty)$, where as by Lemma \ref{aut_synoffrac},  the projective syndromes of  words in the $PGL_2(\f{q})$-orbits of the latter deep hole classes are of the form  $(\mu_a c_{q+1-k}(\mu)+\mu_a^q c_{q+1-k}(\mu^q) )$. Since we have assumed $k \leq q-3$. i.e $q+1-k \geq 4$, it follows that the first two coordinates of  $N_{q+1-k}, N_{q+1-k}+ c_{q+1-k}(\infty)$ are zero. However the 
 first two coordinates of  $(\mu_a c_{q+1-k}(\mu)+\mu_a^q c_{q+1-k}(\mu^q) )$, namely $\mu_a+\mu_a^q, \mu \mu_a+ \mu^q \mu_a^{q}$ are both zero only if $\mu \in \f{q}$ which is not the case. \\

\emph{Remark}:  Consider the standard normal rational curve of degree $(q-k)$ in projective space $\bP^{q-k}(\f{Q})$ over an extension field $\f{Q}$ of $\f{q}$.
Let $P', P''$ be two distinct points on the curve such that not both of them have $\f{q}$-coordinates. Let $P \notin \{P',P''\}$ be a point on the secant line joining $P', P''$. Since any $(q+1-k)$ columns of the matrix $G_{q+1-k}(\f{Q})$ are linearly independent over $\f{Q}$, it follows in particular, that $P$ cannot be written as a $\f{q}$-linear combination of $(q-k-1)$ columns of $G_{q+1-k}(\f{q})$. Thus $P \in \mathbb S(k)$ provided $P$ has $\f{q}$-coordinates. For  $P$ to have  $\f{q}$-coordinates,  $Q$ must be an even power of $q$ so that there is a quadratic extension $\f{q^2}$ of $\f{q}$ in $\f{Q}$, and we must have $P''=\sigma(P')$ where $\sigma$ is the nontrivial automorphism of $\f{q^2}$ over $\f{q}$. Thus the \emph{geometric interpretation} of the $(q+1)q(q-1)/2$ syndromes of the deep hole classes $u(a, P(X))$ is as  follows: there are $(q^2-q)/2$ pairs of distinct points $\{P', \sigma(P')\}$ on the curve, and on the secant line joining $P', \sigma(P')$, there are $(q+1)$ points with $\f{q}$ coordinates.

\section{Complete deep holes of $PRS(q-3)$ }\label{section:k=q-3}
In this section, we will show that the  deep holes constructed in Theorems~\ref{deephole:degreek},~\ref{thm:Rotationsofdegk} and~\ref{thm:main_irr2} form all the deep holes of $PRS(q-3)$.  Since $PRS(k)$ for $k=1$ has been treated in \S\ref{sec_cov_PRS}, we assume $q \geq 5$.  Since Conjecture \ref{cover} is true for $PRS(k)$ when $k \geq \lfloor (q-1)/2 \rfloor$, it follows that  $\rho(PRS(q-3))=3$ provided  $q-3 \geq \lfloor  (q-1)/2 \rfloor$ i.e.  $q \geq 4$, which is the case here.
For $PRS(k)$, even the problem of just determining the number of deep hole classes (not necessarily determining all of them) is very difficult for $k<q-3$. If $\rho(PRS(k))=q-k$, this reduces  to the problem of determining the number of points of $\bP^{q-k}(\f{q})$ which are not in the span of any $q-k-1$ columns of $G_{q+1-k}$. For $k=q-3$, we can calculate this number:
	\begin{thm}\label{numberofdeepholes:k=q-3}
There are $(q^3+2q^2+q)/2$ classes of deep holes of $PRS(q-3)$.
	\end{thm}
	\begin{IEEEproof}
	  The number of deep hole classes of $PRS(q-3)$ is 
		\[ (1+q+q^2+q^3) -  |\{ v \in \bP^3(\f{q}) : v \text{ is in the span of some $2$ columns of $G_4$} \}|. \]
		There are $(q+1)$ points in  $\bP^3(\f{q})$ which are in the span of less than two  columns of $G_4$ (namely $\{c_4(t) : t \in \f{q} \cup \infty\}$). For each of the $\textstyle\binom{q+1}{2}$  pairs of columns $G_4$, there are $q-1$ points which are in the span of these two columns but are not in $\{c_4(t) : t \in \f{q} \cup \infty\}$.	Since any $4$ columns of $G_4$  are linearly independent, 
		a point of $\bP^3(\f{q})$ which is not in $\{c_4(t) : t \in \f{q} \cup \infty\}$, cannot be in the span of two different pairs of columns of $G_4$. Therefore, the number of deep hole classes of $PRS(q-3)$ is 
		\[(1+q+q^2+q^3)-(1+q)-\textstyle\binom{q+1}{2}(q-1) =(q^3+2q^2+q)/2.		\]
	\end{IEEEproof}
We have  shown in Theorem \ref{thm:Rotationsofdegk} that the $q^2$ deep hole classes constructed in this theorem are distinct from the $q$ deep hole classes constructed in Teorem \ref{deephole:degreek}. We have also shown in Theorem \ref{thm:main_irr2}, that the $(q+1)q(q-1)/2$ deep hole classes constructed in this theorem are distinct from the $q^2+q$ deep hole classes of Theorems~\ref{deephole:degreek} and \ref{thm:Rotationsofdegk}.  Since 
\[ (q^3+2q^2+q)/2 = q + q^2+ (q+1)q(q-1)/2,\]
we conclude that all deep hole classes of $PRS(q-3)$ have been found.\\

\section{Conclusion}
The foremost open problem  about deep holes for Projective Reed-Solomon (PRS) codes, is to determine the  covering radius of these codes -- i.e. to settle Conjecture \ref{cover}, or equivalently Conjecture \ref {cover_rnc}. This is a special and important case of the well known MDS conjecture. For dimensions $k$ in which Conjecture \ref{cover} is known to be true, the next important problem is to determine the  deep holes of the code $PRS(k)$. This is a difficult problem. The oldest known  deep holes of $PRS(k)$ are those generated by the polynomial $X^k$. By applying the full  automorphism group of $PRS(k)$ to these deep holes we obtained in this work the deep holes of Theorems ~\ref{deephole:degreek} and ~\ref{thm:Rotationsofdegk}.  
In Theorem \ref{thm:main_irr2}, we obtained new deep holes of $PRS(k)$ using some words having error distance $2$ from the the $\f{q^2}$-linear code $PRS(q^2-q+k)$.
We determined the number of deep holes of  $PRS(q-3)$ and showed in Theorem \ref{main:k=q-3},  that the above two constructions account for all the deep holes of $PRS(q-3)$. For $k<q-3$ it seems increasingly difficult to enumerate the deep holes of $PRS(k)$. The case $k=q-4$  will be discussed  in a forthcoming work.

\bibliographystyle{IEEETranS}
\bibliography{covering}
\end{document}